# GPUTB: Efficient Machine Learning Tight-Binding Method for Large-Scale Electronic Properties Calculations

Yunlong Wang[1,†], Zhixin Liang[1,†], Chi Ding[1], Junjie Wang[1,*], Zheyong Fan[2], Hui-Tian Wang[1], Dingyu Xing[1], and Jian Sun[1,*]

[1] *National Laboratory of Solid State Microstructures, School of Physics and Collaborative Innovation Center of Advanced Microstructures, Nanjing University, Nanjing, 210093, China*

[2] *College of Physical Science and Technology, Bohai University, Jinzhou, P. R. China.*

**ABSTRACT**

The high computational cost of ab-initio methods limits their application in predicting electronic properties at the device scale. Therefore, an efficient method is needed to map the atomic structure to the electronic structure quickly. Here, we develop GPUTB, a GPU-accelerated tight-binding (TB) machine learning framework. GPUTB employs atomic environment descriptors, enabling the model parameters to incorporate environmental dependence. This allows the model to transfer to different basis, xc-functionals, and allotropes easily. Combined with the linear scaling quantum transport method, we have calculated the electronic density of states for up to 100 million atoms in pristine graphene. Trained on finite-temperature structures, the model can be easily extended to millions of atom finite-temperature systems. Furthermore, GPUTB can also successfully describe h-BN/graphene heterojunction systems, demonstrating its capability to handle complex material with high precision. We accurately reproduce the relationship between carrier concentration and room temperature mobility in graphene to verify the framework's accuracy. Therefore, our GPUTB framework presents a delicate balance between computational accuracy and efficiency, providing a powerful computational tool for investing electronic properties for large systems with millions of atoms.

# 1. Introduction

Density functional theory (DFT) methods achieved great success and are widely used in computational materials science; however, they still face challenges in accounting for ionic perturbations on electronic structures at finite temperatures while maintaining computational efficiency. Moreover, such problems are often accompanied by scaling effects, making accurate calculations of electronic properties for systems with hundreds or more atoms computationally expensive [1,2]. Using the linear combination of atomic orbitals (LCAO) basis provides higher computational efficiency than the plane wave (PW) basis [3,4]; however, it remains challenging to directly compute the electronic properties of systems containing millions of atoms with *ab-initio* precision. It is necessary to develop methods that can extend the accuracy of *ab-initio* calculations to a large scale to solve electronic properties calculation at finite-temperature [5]. This is of great significance for promoting the development of multiscale computing and guiding experimental research and applications.

Tight-binding (TB) methods typically employ the LCAO basis, where electronic wave functions are approximated as linear combinations of localized atomic orbitals. This approach allows for an efficient and intuitive description of electronic properties, significantly reducing computational demands and making it suitable for accurately calculating large-scale systems. However, most computational methods to get Hamiltonians based on LCAO basis sets require the explicit construction of overlap matrices [6–8], which incurs additional computational costs. The expensive orthogonalization process makes its application in large-scale electronic properties calculations difficult. Fortunately, the linear-scaling quantum transport (LSQT) method [9–13] uses an orthogonal sparse Hamiltonian, reducing the algorithmic complexity to *O(N)*, which represents a significant achievement in the numerical implementation of quantum transport research over the past few decades and can be used to study experimental scale system, with good application prospects [14,15].

However, the widely used empirical parameter fitting methods for obtaining Hamiltonians are relatively simple and often have limited accuracy [16,17]. Although fitting parameters based on *ab-initio* calculations partially alleviates the accuracy issue, it does not entirely resolve the problem [18–20]. Emerging machine learning techniques have been applied to construct LCAO Hamiltonians. Recently, many E(3) equivariant message-passing neural networks have been proposed to directly learn the Hamiltonian on an LCAO basis [21–23]. Despite achieving excellent accuracy, these methods using a non-orthogonal LCAO basis require orthogonalization calculation. Besides, the large-scale electronic properties methods for non-orthogonal basis sets remain relatively underdeveloped, so extending these methods to large-scale systems is very difficult. Mapping the *ab-initio* Hamiltonian onto the Wannier representation provides an orthogonal Hamiltonian method [24,25]. However, this method constructs a Hamiltonian without considering environmental dependency descriptions, limiting its generalizability to similar systems. Furthermore, it does not avoid the computational processes associated with *ab-initio* calculations and orbital projections. Similarly, TBworks [26] uses the density of states (DOS) as a training target, which effectively addresses the fitting of SK parameters in one-dimensional chain systems but is challenging to generalize to periodic systems. DeePTB [27] uses atomic coordinates and atomic type information to input a multi-layer perceptron (MLP) to obtain environmental descriptors and correct Slater-Koster (SK) parameters by using environment information. By incorporating neural networks, DeePTB has achieved relatively high accuracy. However, graph neural networks (GNNs), being inherently well-suited for handling graph-structured data, often provide more efficient and accurate representations of crystal structures, which naturally exhibit topological relationships, than convolutional networks.

In this work, we developed GPUTB, a machine-learning-based tight-binding (TB) method capable of accurately reproducing ab-initio precision in large systems with millions of atoms. We project the atomic information into a high-dimensional space using Chebyshev polynomial expansion as input to an atomic descriptor network. We also use an environment learning network to introduce environmental dependence. This network exhibits strong band-fitting capability and demonstrates generalizability to similar atomic environments. It can describe allotropes with very different structures (e.g., graphene and diamond) using one single model and heterojunctions (e.g., h-BN/graphene) with remarkable accuracy. Moreover, GPUTB can address systems with spin-orbit coupling (SOC) effects. To generate finite-temperature configurations for large systems with millions of atoms, we employ molecular dynamics (MD) simulations driven by the neuroevolution potential (NEP) [28] and construct its sparse Hamiltonian from the TB model. We have calculated the DOS and conductivity of SiGe in single crystal (3 million atoms) and polycrystalline (2.4 million atoms) and successfully calculated the mobility and carrier concentration relationship of graphene (6.5 million atoms) at room temperature (300 K), which agrees with the previously calculated data and experimental measurements [29–31]. Moreover, our method achieves exceptionally high computational efficiency by benefiting from efficient computing based on GPU (LSQT is implemented in pure CUDA-based code, while the other parts of the code are PyTorch-based), enabling us to calculate graphene's DOS with over 100 million atoms. Therefore, our GPUTB can effectively predict electronic properties at the device scale under finite-temperature conditions, overcoming the long-standing challenge of achieving accurate electronic calculations in such environments.

## 2. Method

### 2.1. Training details

The training process employs the AdamW optimizer and batch normalization to correct gradients and L2 regularization to prevent overfitting. Starting with the lowest energy bands, after ensuring that the MAE change of the previous model within 50 epoch is less than 1%, it is considered fully converged. The converged model is gradually extended to a higher energy band using the same training set. For typical temperature-perturbed structures, this convergence is usually reached in 200-400 epoch (Each epoch represents one step of training for all structures) and 2000~4000 epoch for the 0K structure. Building on this foundation, finite-temperature configurations are incorporated into the training process. Once convergence is achieved, the model can accurately describe thermally fluctuating structures in various environments, enabling application to larger-scale disordered systems with fluctuating configurations. The training equipment consisted of dual Intel Xeon Platinum 8358 CPUs, an A100 SXM4 40GB GPU, and an H100 SXM5 80GB for 100 million atoms. The initial model training began with balanced structures, using a two-atom diamond as an example. The training first focused on the two lowest-energy valence bands, then extended to all four. Subsequently, the fifth, sixth, and seventh conduction bands were gradually added, with each step fully converging before proceeding to the next. After the initial model was trained, it could be rapidly expanded to supercells or other XC functionals.

## 2.2. Finite-temperature molecular dynamics

Training finite-temperature configurations involved ab-initio molecular dynamics simulations on the NVT ensemble using the ABACUS package [8,40,41]. The Nose-Hoover thermostat was set to the corresponding temperature for all configurations, with a time step of 1 fs for 100 ps. From the last 50 ps, 400 structures were selected at equal intervals, with 200 frames chosen randomly as the training set and another 200 as the test set. For large-scale electronic structure calculations, the GPUMD software package [42], based on the NEP force field [43], was employed on the NVT ensemble in conjunction with the Nose-Hoover thermostat, with a time step of 0.5 fs for a total duration of 100 ps. A single structure was randomly selected from the last 50 ps for electronic property calculations. The NEP force field was trained at the target temperature, achieving a root-mean-square error (RMSE) of 3.24 meV/atom for energy predictions and 89.88 meV/Å for force predictions. Molecular dynamics calculations utilized single-crystal structures of 3 million atoms and polycrystalline structures of 2.4 million atoms. Please refer to the Supplemental Material [35] about the NEP force field and GPUMD software.

## 2.3. Ab-initio method

The band structure training set calculations were performed using the *ab-initio* ABACUS package, employing LCAO basis sets in conjunction with SG15 ONCV pseudopotentials for DZP orbitals [44]. The PBE functional was utilized, with additional tests conducted on SCAN and HSE functionals using PW basis sets (Fig. S3). An energy cutoff of 100 Ry was applied, with a minimum allowed spacing of 0.1/bohr between k-points and a charge density convergence threshold of $10^{-8}$ Ry. High-symmetry k-path selections were made for band structure calculations, and in post-processing, targeted bands from the calculations were selected as the training set, disregarding other bands. The NEP force field training set was generated using DFT calculations from the ABACUS package, employing PW basis sets combined with the PBE functional. An energy cutoff of 100 Ry was used, with a minimum allowed spacing of 0.16/bohr between k-points and a charge density convergence threshold of $10^{-6}$ Ry.

## 2.4. Linear scaling quantum transport method

The linear-scaling quantum transport method (LSQT) [10,12,13] is an effective approach for studying electronic transport in large, disordered systems. It incorporates electron-phonon coupling effects through the electronic Hamiltonian, with a computational complexity of O(N), where the cost scales linearly with the number of atoms N. LSQT employs linear scaling techniques, orthogonal polynomials (Chebyshev expansion), the kernel polynomial method (KPM), sparse matrix-vector multiplication, and the random phase approximation to enhance overall computational efficiency [12]. Fan et al. initially developed the CUDA code [11], and we optimized its efficiency and integrated it into GPUTB. The core computational process is as follows:

1. To achieve equilibrium, begin with the initial structure and perform molecular dynamics (MD) simulations over multiple steps in the NVT ensemble.
2. LSQT calculations are performed on multiple frames for systems with significant thermal fluctuations, averaging the results to eliminate thermal fluctuations. For large systems where thermal fluctuations are less pronounced post-equilibration, a single structure suffices for LSQT calculations.

The present work adopts the velocity autocorrelation method to compute conductivity with a time step of 0.1 fs. In this method, the conductivity corresponding to energy $E$ at correlation time t can be expressed as:

$$\sigma(E, t) = \frac{2e^2}{\Omega} \int_0^t \mathrm{Tr}\left[\delta(E - \hat{H}) \, \mathrm{Re}\left(\hat{V}\hat{V}(\tau)\right)\right] \mathrm{d}\tau \qquad (3)$$

Here, $e$ represents the elementary charge, $\Omega$ denotes the volume, $\hat{H}$ is the electronic Hamiltonian operator, $\delta(E - \hat{H})$ is the energy resolution operator, and $\hat{V}$ is the velocity operator. $\hat{V}(\tau) = e^{i\hat{H}\tau}\hat{V}e^{-i\hat{H}\tau}$ gives the time-evolved velocity operator. The DOS is expressed as:

$$\rho(E) = \frac{2}{\Omega}\text{Tr}[\delta(E - \hat{H})] \tag{4}$$

The random states were propagated for over 200 steps to achieve good convergence.

## 3. Result and Discussion

In Fig. 1, we illustrate the workflow of GPUTB. The framework uses a message-passing graph neural network (MPNN) [32,33]. Initially, the input features for the embedded network are generated based on trained crystal structures through element type classification and Chebyshev radial function expansion. This process outputs edge features $e_{ij}$ along with node features $v_i$ and $v_j$. The node features are then fed into an environment network to produce environment-specific features multiplied by the edge features. This product serves as input to the environment-dependent SK parameter neural network, generating Slater-Koster (SK) parameters that adapt to the local atomic environment. These parameters are used to construct the Hamiltonian matrix elements $H_{ij}$. Simultaneously, the node features $v_i$ are processed through a node network to derive the diagonal elements $H_{ij}$ of the Hamiltonian matrix. A loss function is computed by comparing these predictions with reference data from DFT calculations, enabling gradient-based optimization of the model parameters until convergence. Based on this initial 0K model, we introduced all the ion perturbation structures at once to improve the model's generalization ability to perturbations. Finally, the trained model is applied to constructe large perturbed systems' Hamiltonian, and electronic transport properties are obtained by using LSQT calculations.

The TB model assumes that electrons are localized near atoms, with electronic transitions occurring only between neighboring atoms. A set of empirical or semi-empirical parameters is used to describe the hopping integrals between different atomic orbitals, combined with onsite energies to constitute the total Hamiltonian, with the general form being:

$$H = \sum_i \varepsilon_i c_i^\dagger c_i + \sum_{j \in N(i)} t_{ij} c_i^\dagger c_j \quad (1)$$

In this formula, $\varepsilon_i$ represents the onsite energy of a specific orbital on the $i$ atom; the creation and annihilation operators on the $i$ atom are denoted by $c_i^\dagger$ and $c_j$, respectively. $t_{ij}$ is the hopping integral between the $i$ and $j$ atoms, describing the probability of electron transitions from the orbital of $i$ to the orbital of $j$. $N(i)$ is the neighbor atom for atom $i$. Based on the relative positions of the two atoms and their specific orbitals, the corresponding hopping parameters are determined using the SK bond integral [34]. Then, the hopping integrals are simplified into parameters dependent on spatial directions. In GPUTB, the hopping part is $H_{i,j}^{l_i m_i, l_j m_j} = \sum_\lambda MLP_\lambda(e_{ij}) P_\lambda^{l_i m_i, l_j m_j}(\hat{\mathbf{r}}_{ij})$ and onsite part is $H_{i,i}^{l_i m_i, l_i m_i} = MLP_{l_i,\alpha}(v_i)$, where:

$$e_{ij} = C_n(r_{ij}) MLP_{e\lambda}\left(\sum_{k \in N(i)} C_n(r_{ik}) + \sum_{l \in N(j)} C_n(r_{jl})\right); \quad v_i = \sum_{j \in N(i)} C_n(r_{ij}) \quad (2)$$

The $MLP_\lambda$ is the MLP hopping term parameters network, The directional cosine $\hat{\mathbf{r}}_{ij} = \mathbf{r}_{ij}/r_{ij}$, where $r_{ij}$ is the distance between atoms $i$ and $j$, and $P_\lambda^{l_i m_i, l_j m_j}(\hat{\mathbf{r}}_{ij})$ represents the SK conversion coefficient, with $\lambda$ encompassing bond types. $v_i$ and $e_{ij}$ are the node and edge features, and the $MLP_{l_i,\alpha}$ is a MLP onsite network for $l_i$ orbit and $\alpha$ type atoms. The $MLP_{e\lambda}$ is an MLP environment network that uses the node information of atoms $i$ and $j$ as input, and the output is combined with $e_{ij}$ as the input of the SK parameter network. The $C_n(r_{ij}) = \sum_{k=0}^{N_{basis}} c_{nk}^{ij} f_k(r_{ij})$ is the bond environment descriptor and $f_k(r_{ij}) = \frac{1}{2}\left[T_k\left(2(r_{ij}/r_{cut} - 1)^2 - 1\right) + 1\right] f_c(r_{ij})$. $N_{basis}$ is the number of basis functions of the descriptor, and $T_k$ represents the first-type Chebyshev polynomial of order $k$. $c_{nk}^{ij}$ depends on the types of atoms $i$ and $j$. $r_{cut}$ is the cutoff radius, and $f_c(r_{ij})$ is the smooth cutoff function.

GPUTB originates from atomic structures, expanding atomic edges into Chebyshev polynomials as input descriptors for the hopping network and aggregating all neighboring atoms as input descriptors for the onsite network. The descriptor network consists of an MLP with a SiLU nonlinear activation function. Due to symmetry, when constructing the Hamiltonian, $H_{i,j}^{l_i,l_j} = (-1)^{l_i+l_j} H_{j,i}^{l_j,l_i}$, and the construction of half of the edges can be omitted, where bonds of the same type share a pair of parameters. By classifying keys, GPUTB can efficiently generate environment-dependent Hamiltonians, convert them from real space to k-space, and then diagonalize them to obtain $E_{pre}$. Both the descriptor network and the SK parameter neural network are trained by minimizing the loss function, which is defined as:

$$L = \left[\sum_{b,k} w_b |\Delta_{b,k}|\right]^2 + \sum_{b,k} w_b (\Delta_{b,k})^2 \tag{3}$$

In this context, $\Delta_{b,k} = E_{pre}^{b,k} - E_{ref}^{b,k}$, $E_{pre}^{b,k}$ and $E_{ref}^{b,k}$ represent the predicted and reference energy bands from the training set (for the *b*-th band at the *k*-th point), respectively. The reference energy band eigenvalues are obtained from *ab-initio* calculations, and $w_b$ is the weight for the *b*-th band.

Here, we demonstrate the accuracy and generalization capability of the GPUTB method, using SiGe systems as examples. SiGe is representative of semiconductor compounds with diverse atomic environments, making it an ideal benchmark for evaluating the model's accuracy and generalizability, as shown in Fig. 2. Starting from a cubic conventional cell, we employed an atomic orbital basis set including *s*, *p*, and *d* orbitals, with a cutoff radius of 5 Å to include the third nearest neighbors. The initial model was trained on an 8-atom orthogonal cell through an iterative band-expansion strategy, starting with the second band, then progressively incorporating higher bands (up to the 26th), with full convergence ensured at each step before advancing to the next. Subsequently, finite-temperature structures were generated using a machine learning force field, extending the model to a finite-temperature condition. The model exhibited a prediction error of 13.0 meV for the 8-atom structure. This accuracy was retained as the model was scaled to larger systems with the same energy range; the prediction error is 19.5 meV for the 2×2×2 supercell with 64 atoms and 16.7 meV for the 3×3×3 supercell with 216 atoms. The prediction of the 216-atom system using 8-atom training set shows a slight change in the error distribution due to the size effect, but the MAE is still within a reasonable range. We also calculated the computational efficiency of two basis sets, PW and LCAO, of GPUTB and DFT software. The calculation content is the band eigenvalues of the gamma point. The slopes of GPUTB and LCAO are lower than those of larger scales below a few hundred atoms. This is because the subsequent computational overhead is mainly the Hamiltonian's diagonalization rather than the Hamiltonian's generation. The dotted line represents the computational overhead of only considering the generation of the Hamiltonian. This is one of the reasons why the model can be quickly expanded to large systems. In Fig. 2(e), under identical batch sizes, we concurrently train finite-temperature structures on GPUs using both GPUTB and DeePTB, with a computational time of 100 epochs. Benefiting from an efficient network architecture and parallelized optimization for Hamiltonian generation, GPUTB demonstrates superior computational efficiency while maintaining higher accuracy. This approach enables the generation of high-precision orthogonal Hamiltonians for subsequent property calculations.

To more comprehensively evaluate the fitting capability of the GPUTB model, we compared its performance with the DeePTB method across several classical systems, as summarized in Table 1. The results show that GPUTB achieves higher accuracy using the same training set. In the Supplemental Material [35], we present more extensive tests involving various atomic orbital basis sets, exchange-correlation functionals, numbers of nearest neighbors, transition metal systems, and ternary compounds. We also validate the model using different allotropes to demonstrate the universality of the model parameters across diverse structures. Notably, monolayer graphene and diamond exhibit significant structural differences: graphene is a two-dimensional material with *sp²* hybridization, whereas diamond is a three-dimensional material with *sp³* hybridization. As a result, describing both materials using a single model is challenging. However, the GPUTB model accurately captures these distinct atomic environments, reflecting the underlying physical principles and demonstrating its excellent generalizability.

In recent years, Ben Mahmoud et al. [36] proposed a machine learning DOS method that can obtain the electronic state density of a large system from the structure with high accuracy. The process does not involve the explicit solution of the Hamiltonian matrix, so it has a fast speed. However, DOS implies much electronic information, so this method is limited to DOS prediction and cannot provide additional insights, such as the Hamiltonian matrix, necessary for further electronic property calculations beyond DOS. In contrast, the LSQT method, which is based on the electronic Hamiltonian, effectively addresses this limitation. Leveraging the capability of GPUTB to rapidly construct high-precision Hamiltonians directly from atomic structure information, we can efficiently compute transport properties for systems containing millions of atoms. LSQT is capable of effectively describing systems at finite-temperature. As illustrated in Fig. 3, thermal excitations alter the electronic distribution, facilitating thermal transitions of electrons between multiple energy states compared to the DFT tetrahedron method. This results in a dispersion of the DOS near specific energy levels, smoothing the DOS curve and reducing the sharpness of peaks. We further computed the electronic properties of polycrystalline SiGe equilibrated at 300 K. The perturbed atomic arrangements at grain boundaries are considered defect states in polycrystalline materials, introducing additional electronic states. These states form localized states near the Fermi level, leading to a quasi-continuous DOS within the bandgap. Grain boundaries enhance electron scattering, causing frequent scattering events at the boundaries. This reduces the adequate mobility of electrons and hinders their free motion, ultimately leading to a decrease in conductivity across the entire energy range.

As shown in Fig. 4, we generated the Hamiltonian for a graphene system containing over 100 million atoms (approximately 1 billion orbitals) using an *s*, *p*, and *d*-orbital model with a third-nearest-neighbor cutoff. This system's DOS reaches the micrometer scale. The results closely matched those obtained using the tetrahedron method in DFT, demonstrating the reliability of our model. To adapt the model for finite-temperature environments, we employed molecular dynamics simulations by GPUMD in the NVT ensemble at 300 K to sample a 2×2×1 graphene supercell, and the MAE was 25.8 meV. Based on this model, we calculated the mobility of graphene at 300 K as a function of carrier concentration for a system containing over 6.5 million atoms. At a concentration of $10^{11}$ cm$^{-2}$, the results closely aligned with the experimental data reported by Mayorov et al. [29], and at $10^{12}$ cm$^{-2}$, they were consistent with two experimental samples from Chen et al. [30]. Additionally, our results overlapped significantly with the range reported by Shishir et al. [31], further validating the reliability of the LSQT computational method. Moreover, we also successfully employed GPUTB to describe h-BN/graphene heterojunction systems. The h-BN/graphene heterojunctions exhibit novel electronic properties due to symmetry breaking, including bandgap opening at the Fermi level [37–39]. We successfully reproduced the DOS near the Fermi level. We extended the calculations to systems containing millions of atoms, demonstrating its ability to build realistic device-scale systems' electronic properties.

# 4. Discussion and conclusion

In summary, we developed a GPUTB method integrating Slater-Koster parameters with an environment-dependent neural network, achieving accurate Hamiltonians fitting from *ab-initio* electronic calculations. By employing band structures as training targets, GPUTB avoids limitations tied to specific basis sets or functionals, significantly reducing the computational cost of expensive functionals in large systems. GPUTB can construct Hamiltonians for structures at different scales and temperatures, highlighting the validity of its environmental descriptors. Building on this capability, GPUTB accurately captures ionic fluctuations at finite-temperature. With the LSQT method, GPUTB can efficiently compute electronic properties for large-scale systems with millions of atoms, including finite-temperature perturbed single crystals, polycrystalline materials, and heterojunctions. This demonstrates the exceptional computational efficiency of the framework. Complemented by highly accurate Hamiltonian networks, GPUTB can precisely reproduce the relationship between electronic mobility and carrier concentration of graphene at room temperature and the narrow bandgap obtained from first-principles calculations. These results highlight the accuracy, generality, and generalizability of GPUTB, establishing it as a state-of-the-art tool for simulating electronic properties for device scale systems in an experimental environment. It has excellent potential in complex systems, such as polycrystals and heterojunctions. However, there is still space for improvement in the model. For example, the method is based on the LCAO basis set, the network is relatively simple and has few parameters, and it may not be easy to describe complex systems such as such as metal surface or water-metal interfaces. In future versions, the goal will be to introduce more complex networks and expand the completeness of the basis set.

# Declare of competing interests

The authors declare that they have no known competing financial interests or personal relationships that could have appeared to influence the work reported in this paper.

# Acknowledgements

We gratefully acknowledges the National Natural Science Foundation of China (Grant 12125404, T2495231, 123B2049), the Basic Research Program of Jiangsu (Grant BK20233001, BK20241253), the Jiangsu Funding Program for Excellent Postdoctoral Talent (Grants 2024ZB002 and 2024ZB075), the Postdoctoral Fellowship Program of CPSF (Grant GZC20240695), the AI & AI for Science program of Nanjing University, Artificial Intelligence and Quantum physics (AIQ) program of Nanjing University, and the Fundamental Research Funds for the Central Universities. The calculations were carried out using supercomputers at the High-Performance Computing Center of Collaborative Innovation Center of Advanced Microstructures, the high-performance supercomputing center of Nanjing University.

# Author Contributions Statement

J.S. conceived and designed the work. Y.W. performed primary computations and drafted the manuscript. Y.W. and Z.L. proposed the model framework together, Z.L. helped train machine learning potentials, molecular dynamics simulations. J.W. and C.D. helped with the analysis of the molecular dynamics results. Z.F. participated in the discussion of LSQT method. Y.W., H.T.W., D.X. and J.S. wrote the manuscript. All authors participated in scientific discussions and provided critical feedback on the manuscript.

# Figures captions

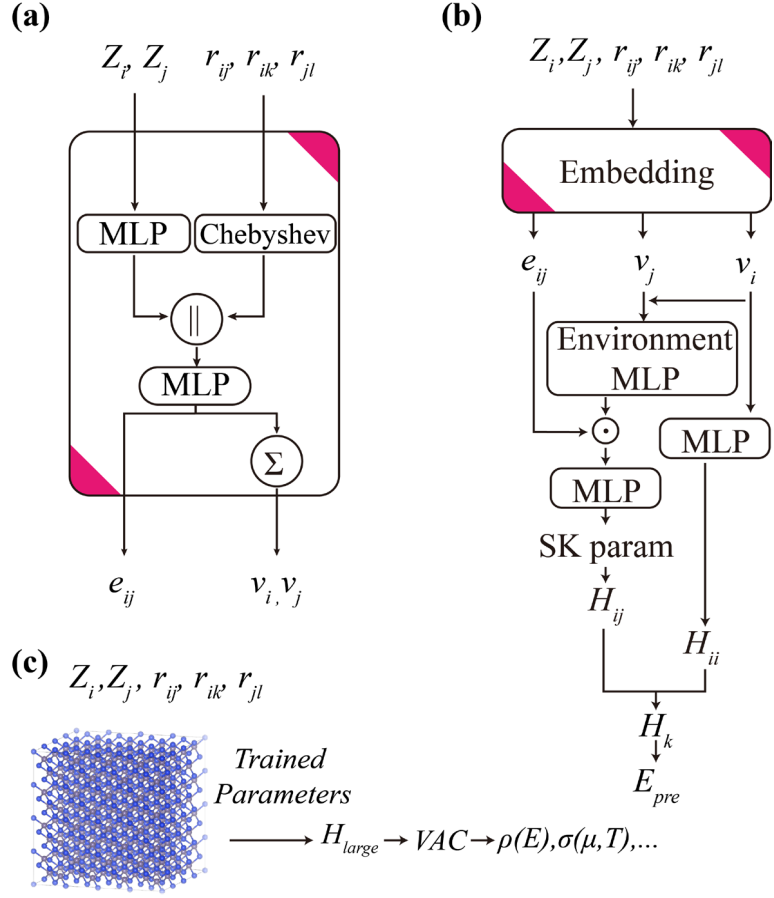

**Fig. 1. Flowchart and network architecture of GPUTB calculations. (a)** The embedding network of GPUTB, where $v_i$, $v_j$ is the node message, $e_{ij}$ is the edge message, $Z_i$, $Z_j$ indicates the types of atoms, and $r_{ij}$, $r_{ik}$, $r_{jk}$ is the distance between neighboring atom pairs ij, ik, jk, respectively. **(b)** Descriptor network and residual parameter network, MLP represents Multi-Layer Perceptron, whereas the Hamiltonian components are represented as $H_{ij}$ (hopping integral), $H_{ii}$ (onsite energy), and $H_k$ (k-space), $E_{pre}$ represents the energy bands predicted by the network. **(c)** The large-scale Hamiltonian $H_{large}$ is constructed by inputting structural information of the macroscopic system into a neural network model. This Hamiltonian is subsequently used to compute the velocity autocorrelation function (VAC), from which critical transport properties, including the density of states (DOS) and electrical conductivity, are derived through post-processing calculations.

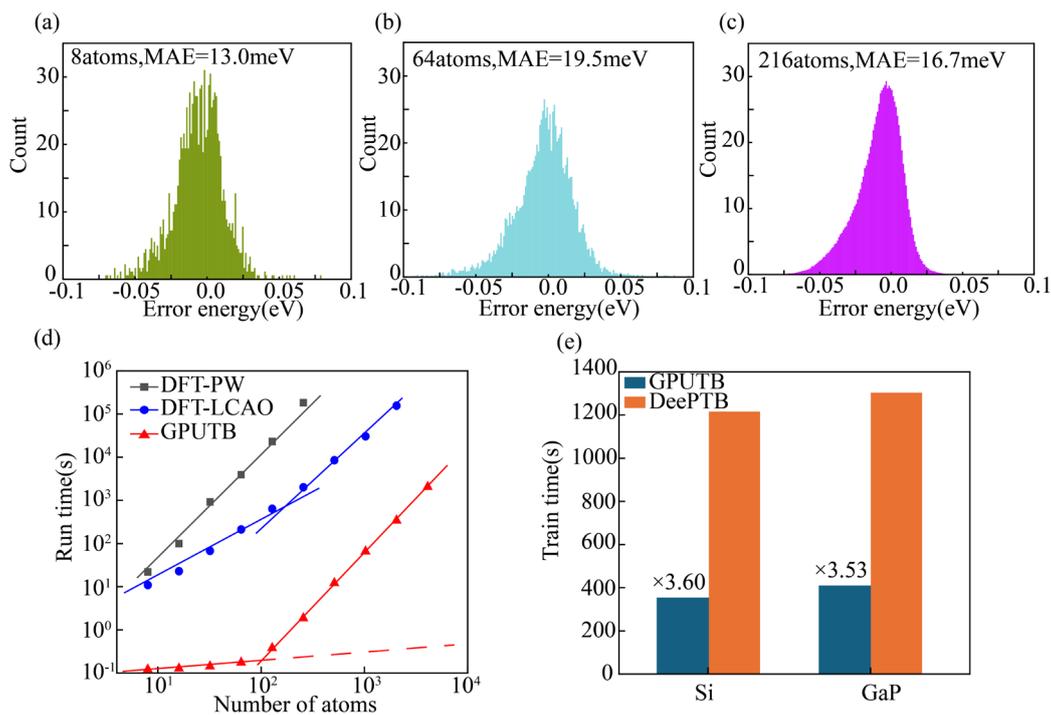

**Fig. 2. Generalization and accuracy of the model at different scales. (a-c)** The error histogram of bands by model trained with 8 atoms predicted for 8 atoms, 64 atoms (2×2×2 supercell), and 216 atoms (3×3×3 supercell) systems compared to the corresponding DFT energies, and the inset shows the energy error distribution. **(d)** Comparison of the time cost of calculating Γ-point eigenvalues with GPUTB, DFT-PW, and DFT-LCAO basis sets. **(e)** The time cost of GPUTB and DeePTB train 10 finite temperature structures simultaneously on GPU with a computation time of 100 epochs.

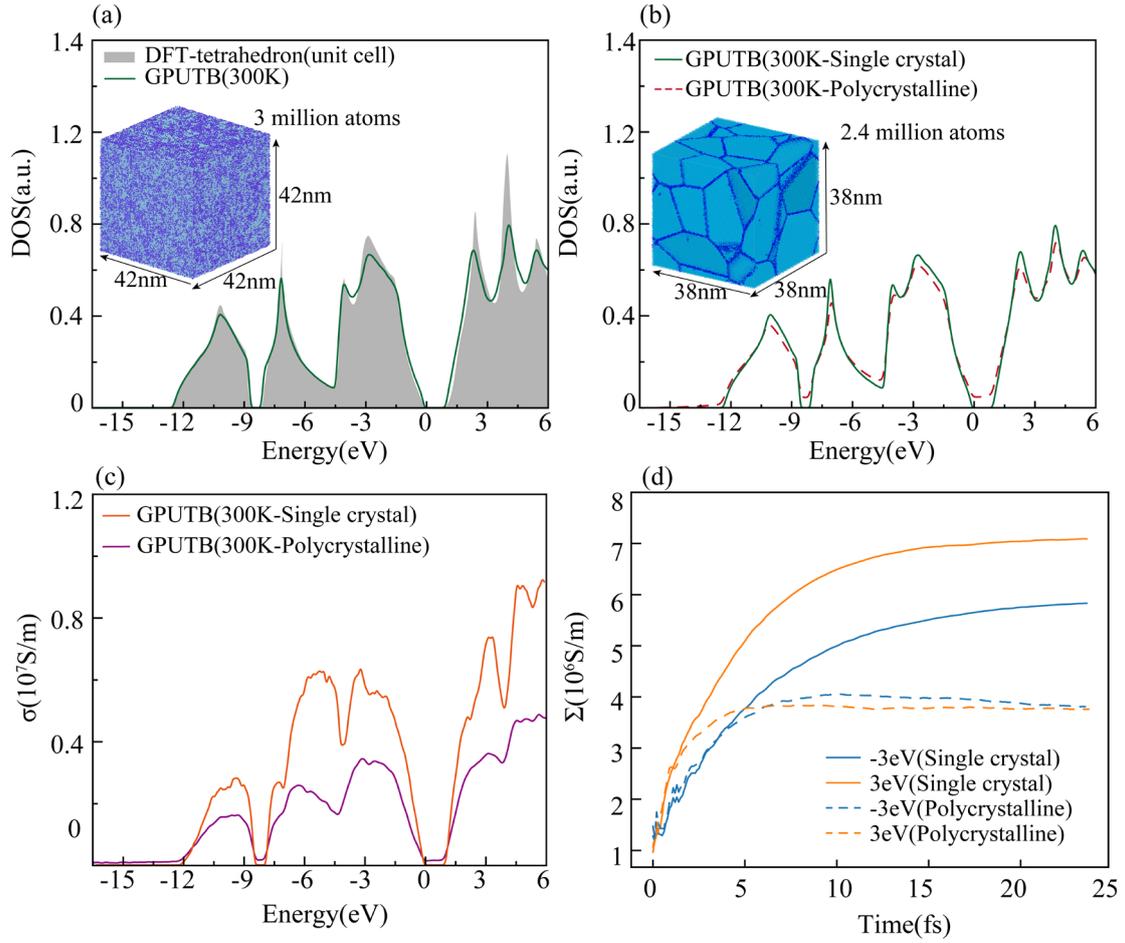

**Fig. 3. Calculate the DOS and conductivity for single-crystal and polycrystalline SiGe at finite temperatures.** **(a)** Comparison of finite-temperature DOS for SiGe at 300K with DFT-calculated DOS for the unit cell at 0 K. **(b)** Comparison of finite-temperature DOS for single-crystal and polycrystalline SiGe. **(c)** Comparison of finite-temperature conductivity for single-crystal and polycrystalline SiGe. **(d)** Evolution of conductivity for single-crystal and polycrystalline SiGe at -3eV and 3eV, converging after 20fs.

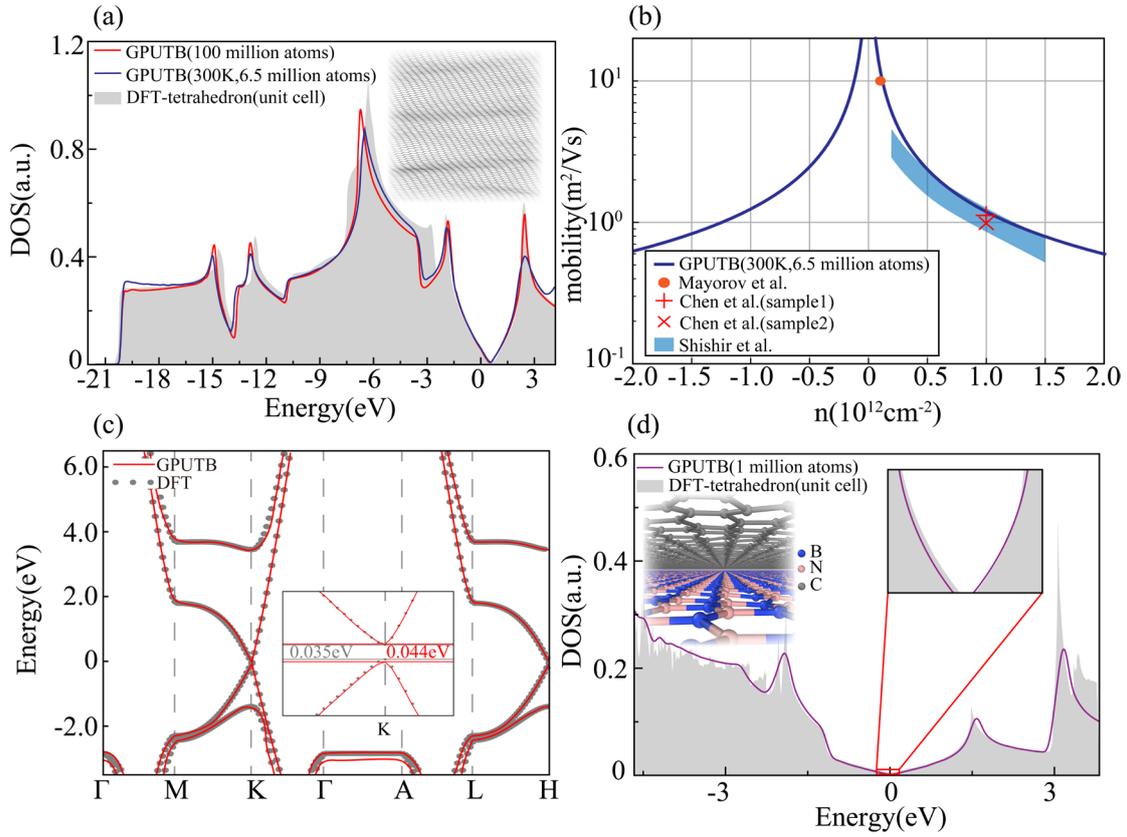

**Fig. 4. Device-scale graphene DOS, carrier mobility, and the DOS of h-BN/graphene heterojunctions. (a)** Comparison of the DOS of graphene containing 100 million atoms at finite-temperature with the DOS computed using the DFT tetrahedron method for a graphene unit cell. **(b)** Carrier mobility of graphene at 300 K as a function of carrier concentration, compared with previous experiments (Mayorov et al. [29] and Chen et al. [30]) and calculations (Shishir et al. [31]). **(c)** Band structure of unit-cell h-BN/graphene heterojunction with 1 million atoms. **(d)** Electronic density of states for a large-scale h-BN/graphene heterojunction system with 1 million atoms.

# Tables

**Table. 1. Comparison of the results between GPUTB and DeePTB** [27]. MAEs are in eV units, c/c represents using cubic phase training and testing, m/c represents using mixed phase training and cubic phase testing, and m/h represents using mixed phase training and hexagonal phase testing.

|  | Diamond | GaP | | | AlAs | | | Average |
| --- | --- | --- | --- | --- | --- | --- | --- | --- |
|  | c/c | c/c | m/c | c/c | c/c | m/c | m/h |  |
| **GPUTB** | **0.016** | **0.012** | **0.012** | **0.016** | **0.020** | **0.021** | **0.035** | **0.018** |
| DeePTB | 0.048 | 0.016 | 0.019 | 0.048 | 0.033 | 0.037 | 0.054 | 0.036 |

# Data availability

Data will be made available on request.